\documentclass[preprint,5p,times,authoryear,twocolumn]{elsarticle}

\usepackage{graphicx}
\usepackage[ansinew]{inputenc}
\usepackage{amsmath}
\usepackage{url}
\usepackage{color}

\journal{Icarus}

\def\mms{$\mathrm{mm\,s^{-1}}$}
\def\cms{$\mathrm{cm\,s^{-1}}$}
\def\metersecond{$\mathrm{m\,s^{-1}}$}
\def\sio{$\mathrm{SiO_2}$}
\def\mum{$\mu$m}
\def\density{$\mathrm{kg\,m^{-3}}$}

\begin{document}

\begin{frontmatter}

\title{Free Collisions in a  Microgravity Many-Particle Experiment. I.\\ Dust Aggregate Sticking at Low Velocities}
\author[igep]{R. Weidling} \ead{r.weidling@tu-braunschweig.de}
\author[igep,kobe]{C. Güttler}
\author[igep]{J. Blum}
\address[igep]{Institut für Geophysik und extraterrestrische Physik, Technische Universität zu Braunschweig,\\Mendelssohnstr. 3, D-38106 Braunschweig, Germany}
\address[kobe]{Department of Earth and Planetary Sciences, Kobe University, 1-1 Rokkodai-cho, Nada-ku, Kobe 657-8501, Japan}

\begin{abstract}
Over the past years the processes involved in the growth of planetesimals have extensively been studied in the laboratory. Based on these experiments, a dust-aggregate collision model was developed upon which computer simulations were based to evaluate how big protoplanetary dust aggregates can grow and to analyze which kinds of collisions are relevant in the solar nebula and are worth further studies in the laboratory. The sticking threshold velocity of millimeter-sized dust aggregates is one such critical value that had so far only theoretically been derived, as the relevant velocities could not be reached in the laboratory. We developed a microgravity experiment that allows us for the first time to study free collisions of mm-sized dust aggregates down to velocities of $\sim 0.1$ \cms\  to assess this part of the protoplanetary dust evolution model. Here, we present the results of 125 free collisions between dust aggregates of 0.5 to 2 mm diameter. Seven collisions with velocities between 0.2 and 3 \cms\ led to sticking, suggesting a transition from perfect sticking to perfect bouncing with a certain sticking probability instead of a sharp velocity threshold. We developed a model to explain the physical processes involved in dust-aggregate sticking, derived dynamical material properties of the dust aggregates from the results of the collisions, and deduced the velocity below which dust aggregates always stick. For millimeter-sized porous dust aggregates this velocity is $8 \cdot 10^{-5}$ \metersecond .
\end{abstract}


\end{frontmatter}

\section{Introduction}\label{sec:intro}
The formation of planetesimals, the precursors of planets, is initiated by the collisional coagulation of small dust particles and aggregates. Velocities are induced from the interaction of these dust particles with the thin gas of the protoplanetary disk (PPD) and the gravitational interaction with the central star \citep{WeidenschillingCuzzi:1993}. For the initially micrometer-sized dust grains, the collision velocities are small enough to let them stick to each other and form larger, fractal dust aggregates \citep{WurmBlum:1998, BlumEtal:2000, BlumWurm:2000, KrauseBlum:2004}. However, as the particles are getting bigger, they decouple more efficiently from the surrounding gas. This leads to an increase in their relative collision velocities \citep{Weidenschilling:1977a} and it is {\it a priori} not clear how large particles can grow by direct sticking. Many laboratory experiments have shown that millimeter-sized particles do not stick to each other if they collide at a velocity which is expected to occur under the conditions in a PPD (see review by \citet{BlumWurm:2008} and recent experiments by \citet{BeitzEtal:2011a}). Indeed, the sticking velocities for those dust-aggregate sizes are so small that they are not yet known and need to be studied, and that is the goal of this paper.

\subsection{The current collision model}\label{sub:intro_model}

Our current knowledge on the outcome of dust-aggregate collisions has been largely shaped by the collision model of \citet{GuettlerEtal:2010}. Based on available results from laboratory experiments, \citeauthor{GuettlerEtal:2010} quantified the outcome of a collision between two dust aggregates in terms of sticking, bouncing, and fragmentation. Moreover, their model predicts which of these outcomes actually occurs for a given set of collision parameters (dust-aggregate masses, dust-aggregate porosities, collision velocity) over a wide range of dust-aggregate masses and collision velocities. According to this model, the sticking velocity for millimeter-sized, porous dust aggregates (i.e. a mass of 0.1 mg) is as small as $10^{-3}$ \metersecond, which is slower than the expected collision velocities for these dust aggregates in a PPD \citep{WeidenschillingCuzzi:1993}. The model of \citeauthor{GuettlerEtal:2010} was implemented into a local growth simulation \citep{ZsomEtal:2010}, and the result was that the growth in a minimum mass solar nebula stalls at masses of approximately 1 mg. Instead of further sticking to each other, the dust aggregates rebound and are compacted as observed in the experiments of \citet{WeidlingEtal:2009}. The picture changed when \citet{ZsomEtal:preprint} included the vertical dimension of the disk and sedimentation of the dust aggregates, because particles with different growth timescales get turbulently mixed and the size distribution becomes wider all over the disk.

Those results are expected to be very sensitive to the exact sticking threshold velocity. In the model by \citeauthor{GuettlerEtal:2010} this velocity is merely based on theoretical assumptions as there had been no experiments for millimeter-sized dust aggregates at the relevant velocities at that time. Therefore, it is most desirable to directly measure the threshold velocity for small, porous dust aggregates and to understand the physical processes involved. An implicit assumption of the \citeauthor{GuettlerEtal:2010} model is that the dust aggregates are homogeneous (but porous) spheres. This means that the model is not necessarily useable for fractal dust aggregates. An exact value for the fractal dimension of dust aggregates in this evolutionary phase remains highly speculative at this point \citep{OrmelEtal:2007, SuyamaEtal:2008, OkuzumiEtal:2009} but it is likely that dust aggregates crossing the threshold from sticking to bouncing for the first time possess a fractal dimension less than three. A measurement of the sticking threshold for fractal dust aggregates of the same size would be an ambitious next step but is not within the scope of this paper.

An additional challenge in this context are the differing existing definitions for fractal dust aggregates, especially for the radius and the volume. \citet{Jones:2011} recently presented a model parameterizing fractal and porous particles made up of finite-sized constituents. He describes such particles by means of an inflation and a dimensionality, which are derived from unambiguous properties of the particles, like the largest spatial extent of the particle or the volume the solid matter occupies. An advantage of this model is that particles do not have to be fractal in the strict mathematical sense to be described with these parameters.

To complement the picture on collisional grain growth in protoplanetary disks, we would like to mention that other effects are considered to play a role but are so far only sparsely studied. These involve magnetic fields and magnetized dust aggregates \citep{NueboldGlassmeier:2000, DominikNuebold:2002} as well as  electrostatic effects to enhance the cross-sectional area and the sticking efficiency \citep{IvlevEtal:2002} and leading to reaccretion of bouncing dust grains \citep{Blum:2004}. Additionally, gas effects might lead to the reaccretion of small fragments after a fragmenting collision \citep{WurmEtal:2001b, WurmEtal:2001a}. Another effect that increases the complexity is the rotation of the dust aggregates. This does not only lead to a more complicated treatment of the velocity (translational and rotational) in each individual collision with a possible influence on the sticking efficiency, but may also change the resulting shape of dust aggregates in the first growth phase \citep{PaszunDominik:2006}.

\subsection{Concept and background of our experiment}\label{sub:experiment_concept}

In order to investigate the transition from sticking to bouncing collisions for mm-sized porous dust aggregates, the particles have to collide with velocities of millimeters per second. With currently available techniques, these velocities are not feasible under standard laboratory conditions, whereas they can be achieved in microgravity experiments. Previous experiments by \citet{HeisselmannEtal:2010} describe a method to achieve very low collision velocities with solid particles: they injected an ensemble of 1 cm diameter glass beads from two opposing sides into a flat box under microgravity conditions. In this granular-gas experiment, each inelastic collision between two particles resulted in the dissipation of energy and, thus, in a lowering of the kinetic energy and velocity. Each additional collision slowed down the particles further, lowering the average collision velocity over time.

For dust particles, the coefficient of restitution -- the ratio of the relative velocity of two colliding particles after and before a collision -- is around $\varepsilon \simeq 0.2$ \citep{BlumMuench:1993, HeisselmannEtal:2007}. Compared to the average coefficient of restitution of the glass beads of 0.64 \citep{HeisselmannEtal:2010} this means that an ensemble of dust particles is slowed down even more efficiently, providing that the average collision frequency is similar. This is evident from the mean velocity of an ideal system of equal-sized particles with velocity-independent $\varepsilon$:
\begin{equation}
    v(t) = \left\{ \frac{1}{v_0} + \left(1-\varepsilon\right) n \sigma t\right\}^{-1}\ , \label{eq:haff}
\end{equation}
\citep{HeisselmannEtal:2010} generally referred to as Haff's law \citep{Haff:1983}. Here, $v_0$ is the initial particle velocity at time $t=0$, and $n$ and $\sigma$ are the number density and the collision cross section of the particles.

In order to be able to follow the trajectory of each particle throughout the whole experiment, \citeauthor{HeisselmannEtal:2010} chose the dimensions of the chamber containing the glass beads to be 1.5 times as high as a single sphere diameter, preventing them from obscuring each other. A side effect of this quasi two-dimensional setup is that the particles collide with the walls very often. In the case of glass particles and glass walls this effect can be neglected (the coefficient of restitution in particle-wall collisions in their case was significantly higher than in particle-particle collisions). However, porous dust particles tend to stick to glass walls even at moderate velocities, which requires a larger test volume compared to the particle size. This leads to a slightly different implementation of the concept outlined by \citeauthor{HeisselmannEtal:2010} and the details of our setup are described in the next section.

\section{Experimental Setup}\label{sec:experimental_setup}
Our experimental setups are described in detail in Sects. \ref{sub:setup_medea1}, \ref{sub:setup_medea2}, and \ref{sub:setup_medea3}. To account for technical advances in the past and future, we will refer to the experiment by the acronym MEDEA (Microgravity Experiment on Dust Environments in Astrophysics) together with a number for the respective version of the experiment. In future efforts, the improvements of the experimental setup will help us to address some issues more accurately than we are able to do in this work.

\begin{figure*}[t]
    \includegraphics[width=\textwidth]{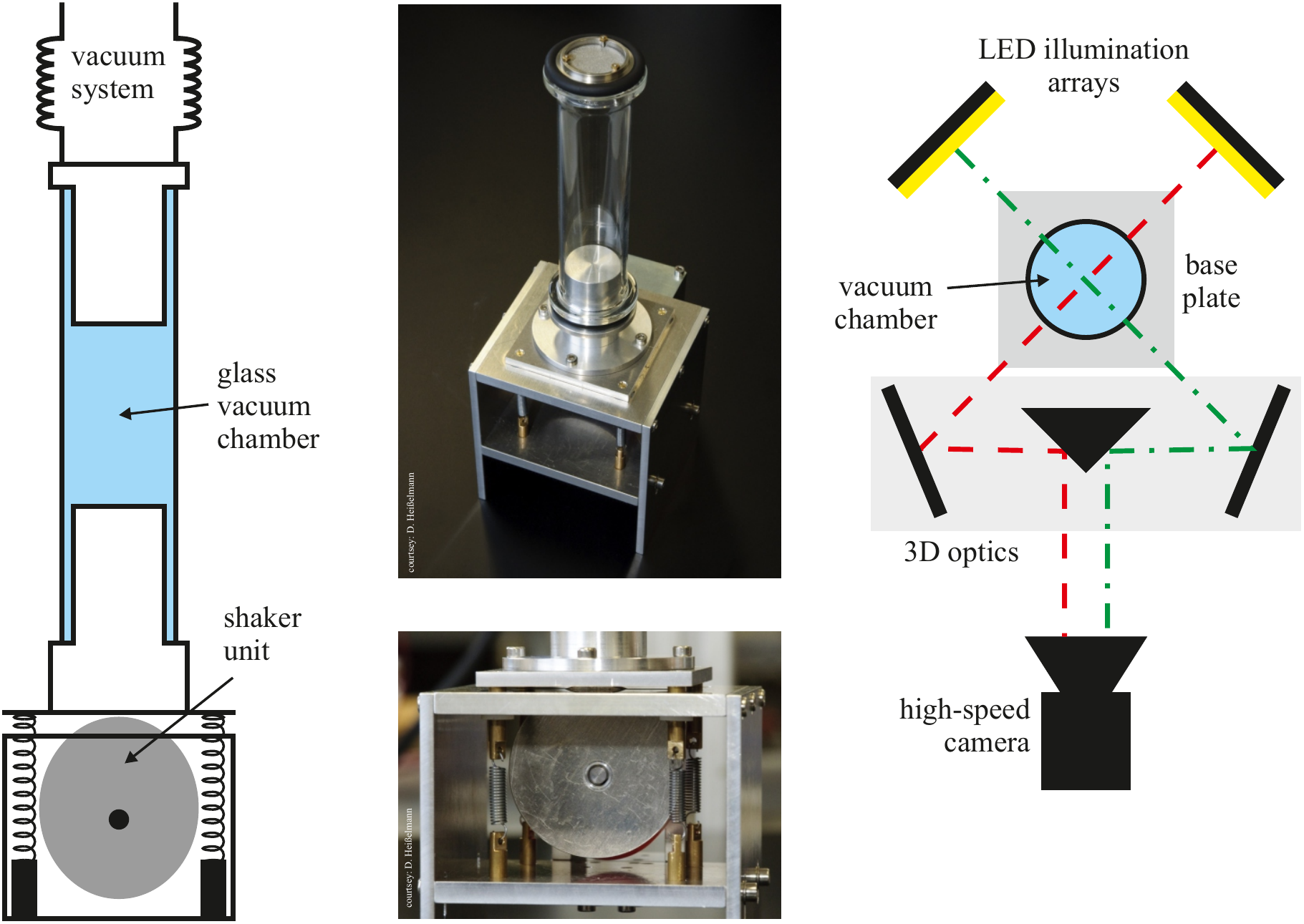}
    \caption{(color online) \emph{Left:} The MEDEA setup consists of a shaking mechanism with an eccentric wheel and a glass vacuum chamber, which is connected to the vacuum system. \emph{Middle:} Photograph of the setup and a close-up of the shaking mechanism. \emph{Right:} Optical system of the MEDEA-I experiment with high-speed camera, beam splitter, and back-light illumination. The red dashed line and green dash-dotted line illustrate the optical paths.}
    \label{fig:sketch_shaker}
\end{figure*}

In contrast to \citeauthor{HeisselmannEtal:2010} we chose a cylindrical geometry for our test chambers (see Fig. \ref{fig:sketch_shaker}). We use glass vacuum chambers with a diameter of 25 mm and a height of 50 mm. These can be agitated in a sinusoidal oscillation along the cylinder axis to excite the dust particles inside. In microgravity, the dust aggregates are observed with high-speed cameras in back-light illumination. According to Equation (\ref{eq:haff}), a high number density of dust particles is desirable, which however constrains the observability of the dust aggregates. To optimize collision time and observability, we chose an optical depth of approximately 0.3, resulting in number densities from $n = 5 \cdot 10^5$ to $5\cdot 10^9$ m$^{-3}$ depending on the particle size. For 0.75 mm diameter dust aggregates this leads to a collision time $\tau = \left( n\sigma v\right) ^{-1} \approx 0.16$ s, with $v = 0.1$ \metersecond\ being a typical relative velocity at the beginning of the experiment.

\subsection{First Setup (MEDEA-I)}\label{sub:setup_medea1}

The first experimental setup, which we will refer to as MEDEA-I, was used in an experiment campaign at the Bremen drop tower in June 2010. Utilizing the catapult, we obtained a microgravity time of nearly 9.5 seconds in each of our 10 flights. We used four identical setups, i.e. four test chambers and four cameras, in each flight, resulting in 40 experiments to extensively test and develop the setup and to vary parameters like dust-aggregate size, dust-aggregate material, number density, agitation frequency, etc.

As mentioned above, each vacuum chamber has a cylindrical shape with a diameter of 25 mm, a height of 50 mm and consists of glass. The bottom flange is made of aluminum and contains a magnetic coil able to inject a permanent magnet into the chamber. The magnet serves as a fast solid particle that can break up agglomerates of dust particles or accelerate a small part of the dust particles in the chamber when colliding with them. The upper flange consists of a metallic grid (mesh size 100 \mum), which allows for the evacuation of the chamber while keeping the dust aggregates in the chamber. On top of each chamber, vacuum bellows provide the connection to the vacuum system. The flanges and the vacuum chamber can be removed from the setup in order to place the particles in the chamber and clean it (Fig. \ref{fig:sketch_shaker}, left and top middle).

Below the vacuum chamber, a shaking mechanism enabled us to excite the vacuum chamber (Fig. \ref{fig:sketch_shaker}, left and bottom middle). The mechanism consists of a DC motor driving an eccentric wheel with a horizontal axis. The bottom of the lower flange is pushed against the top point of the wheel, which causes it to move up- and downwards. Springs between the baseplate and the bottom flange as well as the vacuum bellows ensure the contact with the wheel even in microgravity. By this mechanism, the test chamber was forced to oscillate in an approximately sinusoidal motion with an amplitude of $A = 1$ mm to $A = 2$ mm and an angular frequency between $\omega = 2\pi \cdot 4$ s$^{-1}$ and $\omega = 2\pi \cdot 16$ s$^{-1}$. At the beginning of the experiment, the shaking mechanism distributes the particles in the vacuum chamber, and later it ensures that even in case of residual accelerations that cause particles to drift to the top or bottom, the particles get pushed back to the center of the chamber which is keeping the density homogeneous. According to the amplitude and frequency of the agitated top and bottom flange, collisions with the flanges occurred at a flange velocity of $A\omega=0.02 .. 0.2$ \metersecond\ and an acceleration of $A\omega^2=0.6 .. 20$ $\mathrm{m\,s^{-2}}$.

Each experiment is observed with a high-speed camera operated at 500 frames per second with a resolution of $500\times 500$ pixels. In front of the experiment, a mirror with a $90^{\circ}$-angle facing towards the camera serves as a beam splitter. A mirror at each side reflects the beam towards the vacuum chamber, enabling us to get two rectangular views of the experiment. Behind the chamber, two LED arrays and diffusers are placed to illuminate the experiment (Fig. \ref{fig:sketch_shaker}, right). With this system, the optical resolution of the camera was 120 \mum\ per pixel.

All four experiments were connected to the same vacuum system with a vent line out of the drop capsule. Through this vent line, the vacuum chambers were slowly evacuated, together with the drop tower, down to a final pressure of 10 Pa. The valve was then closed right before firing the catapult and launching the experiment.

\subsection{Second Setup (MEDEA-II)}\label{sub:setup_medea2}

The setup introduced above was used for the results presented in this paper. However, as we also qualitatively checked our results with an improved setup and will present more of these in the future, we will also address the improvements of our setup after the first campaign. We already conducted a second drop tower campaign with a slightly modified setup in November 2010 -- again with four experiment chambers, but this time with a total of 15 catapult flights. We will refer to this setup and experiment campaign as MEDEA-II. In addition to varying particle size, particle density and shaker modes, a broad variety of materials were tested in this campaign (see forthcoming papers).

One important change was to modify the viewing angle of the camera as compared to the first campaign. As our first analysis of the data showed, the two views separated by $90^{\circ}$ are impossible to correlate in this many-particle experiment. Therefore, we reduced the viewing angle to $30^{\circ}$, putting this into practice by substituting the beam splitting mirror in Fig. \ref{fig:sketch_shaker} by a refraction prism ($130^{\circ} \times 25^{\circ} \times 25^{\circ}$) and omitting the two outer mirrors. A view through the prism directly yields two images separated by $30^{\circ}$ and we chose to look onto the wide plane of the prism with the $130^{\circ}$ corner pointing to the experiment chamber. With this flat viewing angle it was sufficient to use only one illumination array per experiment, placed in a line with camera, prism, and vacuum chamber.

To ensure a better performance of the shakers at low frequencies, we installed a servo controller for each DC motor. In contrast to a simple power line, the controllers were able to compensate higher mechanical loads by a higher current. Thus, we could operate the shaking mechanisms at lower frequencies without risking them to jam. This is particularly useful if no residual accelerations cause the particles to drift, but one still wants to have a means of removing them from the top and bottom walls.

When we used the reduced pressure in the drop tower for the MEDEA-I campaign (10 Pa), we found that gas drag still played a minor role for long particle trajectories and for small particles (cf. \ref{app:gasdrag}). To avoid this, we used an additional turbomolecular pump in the MEDEA-II campaign. This was operated while the experiment was prepared to be shot in the catapult before the vacuum was sealed with a valve and the pump flooded. From sealing the vacuum until launching the experiment it took about 30 seconds and we achieved an experiment pressure of 0.1 Pa. Three of our four experiment chambers were connected to this vacuum system while a fourth chamber was not evacuated but operated at ambient pressure ($10^5$ Pa). We used this setup to study free collisions between solid particles, where gas drag plays a negligible role.

\subsection{Third Setup (MEDEA-III)}\label{sub:setup_medea3}
In a third drop tower campaign with 5 catapult flights in August 2011 we mainly used the setup mentioned above with only slight improvements including \emph{(i)} an improved version of the excitation mechanism that is less prone to get stuck, \emph{(ii)} a better illumination system, fitted to the 30° optics, and \emph{(iii)} a restricted experiment volume in some experiments. Fitting the size of the camera images, the insets to reduce this volume had a depth of only a few particle diameters, allowing us to increase the number density of the dust aggregates while keeping the optical depth low enough to observe all the particles. Some of the inner walls of the insets were coated with a nano-material to reduce the effective surface and, thus, to prevent sticking of the dust aggregates to the glass walls. We will refer to the experiments conducted in that campaign as MEDEA-III.

\subsection{Dust analog material}\label{sub:dust}

\begin{figure}[t]
    \includegraphics[width=\columnwidth]{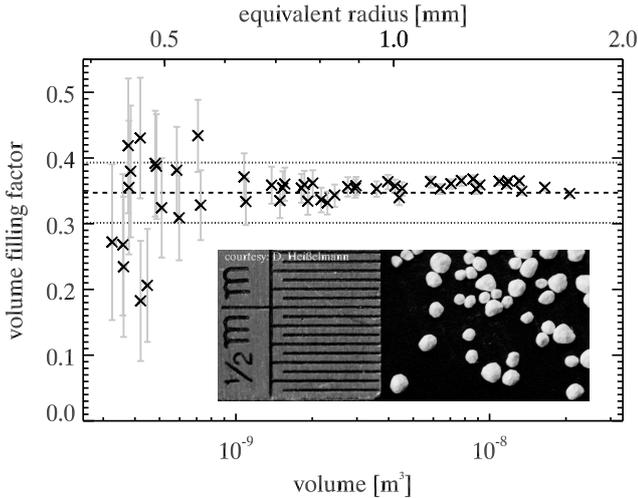}
    \caption{The volume filling factor of the dust aggregates used in the MEDEA-I experiments. On average they possess a filling factor of 0.35, denoted by the dashed line. The gray bars denote the errors due to uncertainties in the measurement of the mass and volume of the particles and the dotted lines show the standard deviation from the mean value. The inset shows a photograph of typical dust particles as used in the experiments described in Sect. \ref{sec:results}.}
    \label{fig:porosity}
\end{figure}

For the experiments described in the following, we used dust particles consisting of irregular-shaped, polydisperse \sio, with $\approx$80\% of the particles having a diameter of 1 to 5 \mum\ ($\approx$99\% between 0.5 and 10 \mum, manufacturer information) and a material density of 2600~\density\ (see Table 1 in \citet{BlumEtal:2006} for more details on its properties). Dust from the storage container forms natural aggregates, which were sifted before the experiments to obtain a narrow dust-aggregate size distribution (see inset in Fig. \ref{fig:porosity}). We used sieves of 0.5, 1.0, and 1.6 mm mesh sizes, which resulted in particles with sizes either between 0.5 and 1.5 mm (henceforth referred to as 1 mm dust aggregates) or between 1.0 and 2.0 mm (hereafter referred to as 1.5 mm dust aggregates). The largest particles can always be slightly bigger than the mesh size due to the fact that the particles are ellipsoids and slip through the grid if their semi-minor axes are smaller than the mesh size.

To measure the volume filling factor $\phi$ (the fraction of the dust-aggregate's volume actually occupied by dust particles), we examined dust aggregates with diameters of about 0.5 to 4 mm from the storage container. We developed a setup in which a particle can be placed on a rotatable platform with a vertical rotation axis. Each individual dust aggregate was observed with a high resolution camera (1280 $\times$ 1024 pixels, 4 $\mu$m per pixel) and back-light illumination. After a full rotation of the particle around one axis with about 100 images taken, we measured the diameter of the particle for each horizontal line in the acquired image sequence. By determining the largest and smallest diameter we then calculated the volume of each slice, assuming an elliptic shape. Adding up these volumes and accounting for shadow effects of the sample holder, this yields the total volume of a particle. Although a polygon would be a better approximation for each horizontal slice, this would overestimate the particles volume if shadowing effects due to concave regions occur. As the particles in general have a convex shape, the assumption of elliptic slices, where only information about the largest and smallest diameter are required, is more accurate in our case.

After the determination of the volume, the dust aggregates with masses ranging from 0.1 to 30 mg were weighed to an accuracy of $0.1$ mg. The results are presented in Fig. \ref{fig:porosity}, where a larger error in the filling factor is evident for the smaller particles, which is mainly due to the uncertainty in the mass determination. In contrast to that, the error in the determination of the volume is very small and can barely be seen. As the measurements for the volume filling factor scatter equally to higher and lower values around the average and all of the dust aggregates are formed in the same way, we assume the same volume filling factor of $\phi = 0.35 \pm 0.05$ (standard deviation) for dust aggregates of all sizes. The shapes of our dust aggregates can also be expressed by an inflation $I$ according to the model of \citet{Jones:2011}. As the dust aggregates consist of monomer particles with a size distribution, which do not occupy positions on a regular grid, we follow his equation (34) and arrive at $I = 2.9$, which is in this case the same as the enlargement parameter defined by \citet{OrmelEtal:2007}.

In order to determine the irregularity of the MEDEA-I particles, we measured the aspect ratio of our particles by analyzing the change of the cross section of rotating particles in the microgravity experiments. If we follow the particle trajectories over a sufficient number of rotations, we can assume to have observed the dust aggregates from all sides and, thus, also observed the largest and smallest axis. Assuming that the intermediate axis is observed in both cases, the ratio of the largest to the smallest cross section equals the ratio of the largest to the smallest axis. In our case, a typical aspect ratio is 1:$x$:1.7, where $1 < x < 1.7$. Using equation (33) of \citet{Jones:2011}, this yields a dimensionality of 2.2 for rod-like particles to 2.6 for disk-like dust aggregates.

\section{Results}\label{sec:results}

We completed the analysis of one experiment with 1.0 mm dust particles and two experiments with 1.5 mm dust particles from the first drop tower campaign in June 2010 with the MEDEA-1 setup (cf. Sect. \ref{sub:setup_medea1}). Other experiments with similar sized dust particles (including the measurement campaigns described in Sect. \ref{sub:setup_medea2} and \ref{sub:setup_medea3}) were so far only visually inspected but qualitatively agree with the results presented here.

In order to analyze the experiments, the images from the high-speed camera were first separated into the two perpendicular views and then the images were binarized. The threshold was selected such that the size (i.e. area) of the particles in the center of the image was reproduced as well as possible in the binary image, as we used the cross section to calculate the particle mass assuming spherical shape and the volume filling factor presented in Sect. \ref{sub:dust}. Due to imperfect illumination, particles at the darker edges appeared slightly larger in the binarized image, but this has no influence on the detection of the center of mass.

In the next step, the position of each particle that did not overlap with another particle and could be distinguished from the background was tracked for as long as it was visible. The position was computed as the center of the area in the binary image and then stored together with the frame number (i.e. time) and the cross sectional area. To achieve statistically unbiased results, we followed every single particle in the experiment with the 1 mm particles consecutively. Every time a particle could unambiguously be identified, it was tracked as long as it could be followed. We then returned to the image where this particle was first tracked and continued to look for the next particle from there. For the larger 1.5 mm dust aggregates, we did not find any sticking event and only aimed at measuring the collision velocities for a representative number of bouncing collisions. We looked for these collisions and only followed the colliding particles before the incident. In the following, we will focus on the analysis of the 1 mm dust aggregates if not stated otherwise.

If a particle track deviated from the ballistic trajectory, it was visually checked whether a collision with another dust particle or the chamber wall had occurred. In the former case, the number of the tracked particle, the time of the collision and the duration of the undisturbed track before and after the collision were noted. Collisions with the wall were ignored, as they slow down the particles but are of no other interest for us. To deduce the particle velocity, we linearly fitted the uninterrupted parts of the tracks. For tracks longer than 60 sequential images, we split the track into segments as a track may not be perfectly linear due to minor gas drag effects (cf. \ref{app:gasdrag}). The velocity of each of these segments was calculated separately for the $x$- (perpendicular to the shaking direction) and for the $z$-component (parallel to the shaking direction), then geometrically added up and assigned to the point in time in the middle of the segment. We did not succeed in correlating the two camera views for maintaining full three-dimensional trajectories. This is why we changed the setup in the MEDEA-II and -III campaign to a $30^\circ$ viewing angle presented in Sect. \ref{sub:setup_medea2}, in which the two images are in a closer relation to each other. In \ref{app:projection} we show that the velocity component parallel to the shaking direction is about twice as fast as the components perpendicular to it. If we choose a $z$-axis along the direction of excitation (perpendicular to the top and bottom flanges), there is no reason why the velocities in the orthogonal $x$- and $y$-directions should be different to each other. This means that the dimension we do not see is comparable to the slower one that we do observe. Adding up the components geometrically this results in an underestimation of the velocity of 13 \% if we only take the value from the projection. Nevertheless, we will use the projected velocity henceforth, as we only know a statistical but not the exact value for the dimension that is not observed.

\begin{figure}[t]
    \includegraphics[width=\columnwidth]{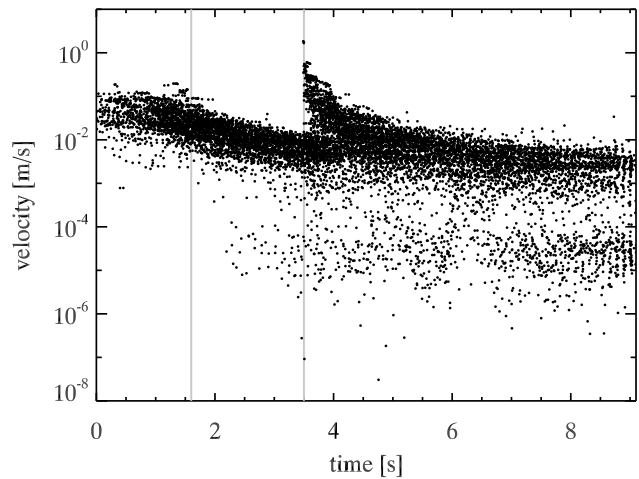}
    \caption{\label{fig:velocitydecay}The average velocity of the 1 mm dust aggregates decreases from about 0.1 \metersecond\ to a few millimeters per second during the experiment. The vertical lines denote the times where the shaker was switched off and then a fast solid object was injected into the experiment chamber.}
\end{figure}

In Fig. \ref{fig:velocitydecay} the absolute two-dimensional velocities of all tracked 1 mm dust aggregates during the experiment are plotted over time. Each dot represents one dust aggregate in one time interval. At the beginning, the particles move with velocities of 0.1 to 0.01 \metersecond, which they achieve from collisions with the top and bottom flange. Two distinct events are marked with vertical lines: after $t = 1.6$ s, the shaker was turned off and the mean velocity immediately decreases due to inelastic collisions of the particles. Although energy was also dissipated before, the shaking induced additional energy which kept the average velocity approximately at a constant level. Some particles are also decelerated in this phase, as after about 0.8 seconds the cloud of particles reaches the upper flange, is reflected and the number of collisions increases dramatically. At the second event, after 3.5 s, a magnet was accelerated into the vacuum chamber (cf. Sect. \ref{sub:setup_medea1}), introducing a fast, solid object colliding with several of the dust aggregates on its way. These particles were accelerated to high velocities, but mutual collisions quickly slowed down the dust aggregates again. Some particles appear to move at velocities between $10^{-6}$ and $10^{-4}$ \metersecond, where the lower value of this range is given by our resolution limit. In reality, all of these seemingly extremely slow particles are sticking to the wall of the vacuum chamber and a residual motion of the chamber or the particles, resulting in slight changes in the binary image of the particles, is then misinterpreted as a very low velocity. As we will in the following only treat free collisions between dust aggregates, these particles are not considered any further. Most of the free-floating particles possessed velocities between $10^{-4}$ and $10^{-2}$ \metersecond\ by the end of the experiment, resulting in collisions with relative velocities also in this range.

It is obvious from the velocity evolution that the dust aggregates collide with one another. However, most of these collisions could not be analyzed as particles may be colliding within dense clusters or only while close to other particles (overlapping in the binary images). However, a total number of 103 collisions between 1 mm dust aggregates could be observed well enough to analyze them in more detail. We do not expect these to be biased in any way and therefore regard them as a representative subset. For the 1.5 mm dust aggregates, we observed 22 representative collisions. Further analysis of these collisions is presented below.

\subsection{Collisional outcomes}\label{sub:collisionresults}

\begin{figure}[t]
    \includegraphics[width=\columnwidth]{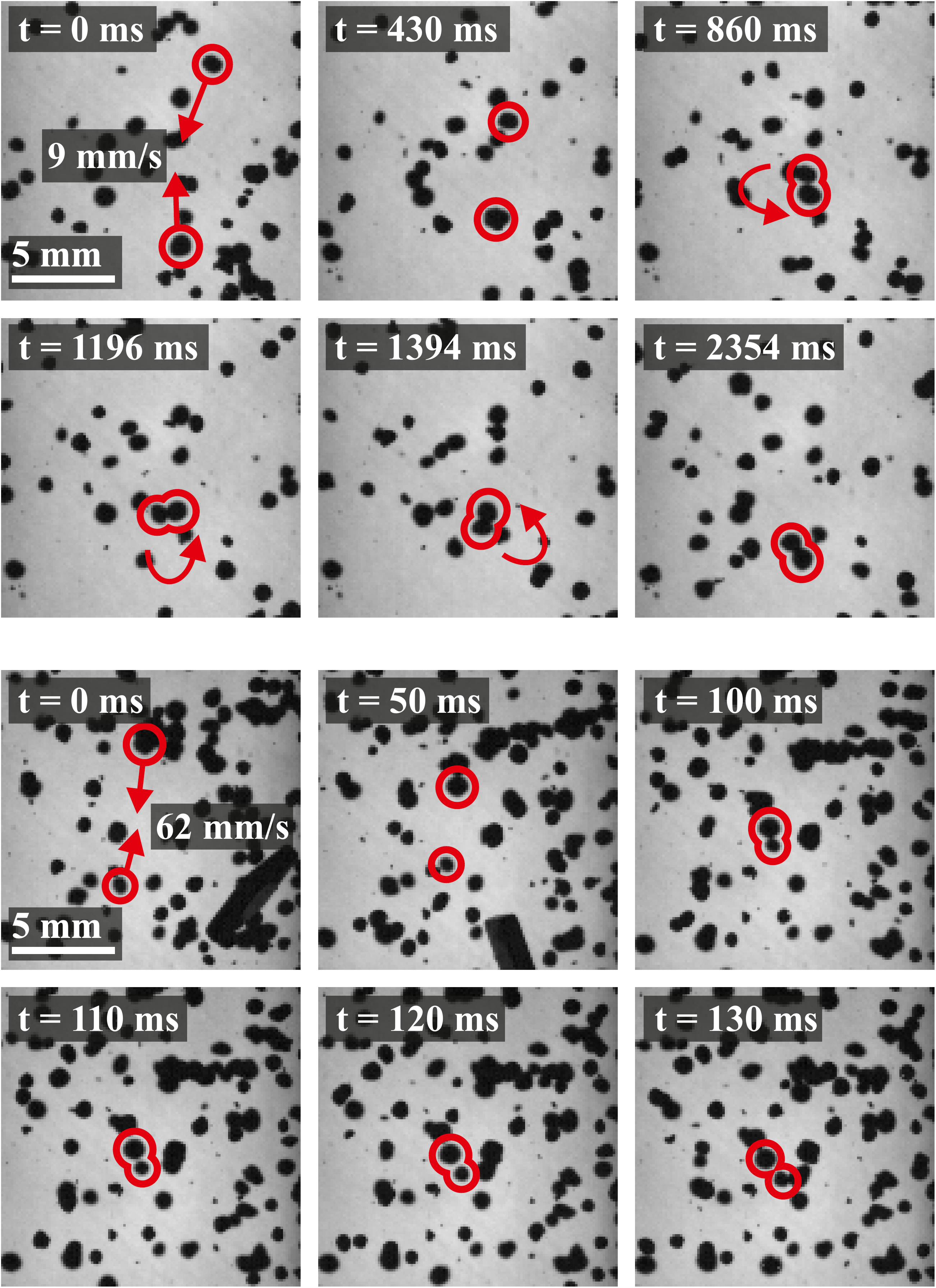}
    \caption{\label{fig:sequence}(color online) \emph{Top:} An image sequence of a sticking collision at a velocity of 9 \mms. After sticking to each other, the particles rotate around their common center of mass. \emph{Bottom:} For higher velocities (62 \mms\ in this case) the collisions often lead to rebound. At $t = $100 ms, the particles are in contact and clearly separate on the subsequent images. The unequal choice of the time steps is due to shadowing effects of other particles in the many-particle system. A movie of this collision can be found in the online version of this article.}
\end{figure}

The most prominent result of each collision is its outcome in terms of sticking, bouncing, and fragmentation. From the 103 analyzed collisions of 1 mm dust aggregates, one resulted in the fragmentation of one of the dust aggregates, seven in coagulation and 95 in bouncing. Two exemplary sequences for a sticking and a bouncing collision are presented in Fig. \ref{fig:sequence}. After a sticking collision, the two particles rotate around their common center of mass and, as a requirement for positive detection of sticking, we demand that this rotation is at least 180 degrees if we do not have any other evidence that they actually stick together, like information from the other viewing direction or a collision effecting both particles (see Sect. \ref{sec:dimer}). In the top part of Fig. \ref{fig:sequence}, two dust aggregates collide with a relative velocity of $9\cdot 10^{-3}$ \metersecond, stick together and are rotating around each other several times. The image sequence in the bottom of Fig. \ref{fig:sequence} shows two dust aggregates colliding with $6.2\cdot 10^{-2}$ \metersecond\ which bounce off after the contact (at $t$ = 100 ms), noticeable by the growing distance between them afterwards. At $t = 0$ and $t = 50$ ms, the injected magnet described in Sect. \ref{sub:setup_medea1} can be seen in the bottom right corner.

All of the 22 collisions of the larger 1.5 mm dust aggregates resulted in bouncing. Although not all particles were tracked in these experiments, the analyzed collisions are a representative subset, as we could not detect any collision resulting in sticking or fragmentation.

\begin{figure}[t]
    \includegraphics[width=\columnwidth]{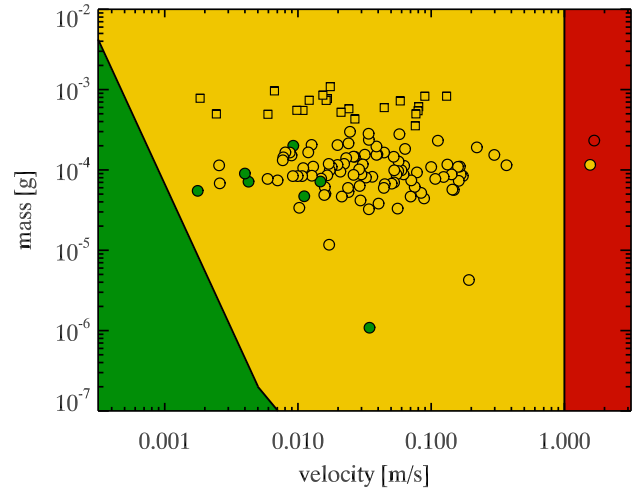}
    \caption{\label{fig:parameterplotwithdata}(color online) In this velocity-mass parameter plot, most of our collisional outcomes agree with the collision types predicted by \citet[][background color]{GuettlerEtal:2010}. Green color denotes sticking collisions, yellow describes bouncing, and red marks fragmentation; the circles represent the results for the 1 mm dust aggregates, the squares those for the 1.5 mm dust aggregates. All of the collisions that led to sticking in the experiment are in disagreement with the model of a sharp transition from sticking to bouncing.}
\end{figure}

In Fig. \ref{fig:parameterplotwithdata}, all analyzed collisions are arranged according to their relative velocity and the mass of the smaller dust aggregate in the collision; the colors denote sticking (green), bouncing (yellow), and fragmentation (red), the symbol distinguishes between the experiment with the small 1 mm dust aggregates (circles) and the two experiments with the larger 1.5 mm particles (squares). The color in the background is the expected collisional outcome following the model of \citet{GuettlerEtal:2010}. The majority of our data matches the model as it predicts bouncing for the velocities we observed in our experiment. One particularly fast collision, triggered by the injection of the magnet, occurred at a relative velocity of 1.7 \metersecond\ and resulted in fragmentation of a particle.

However, we also observed for the first time millimeter-sized dust aggregates stick to one another under controlled experimental conditions. The collision velocity in these cases ranged from $3 \cdot 10^{-2}$ to $2 \cdot 10^{-3}$ \metersecond. This was unexpected as the dust-aggregate collision model predicts sticking of 0.1 mg dust aggregates to occur only at velocities lower than $10^{-3}$ \metersecond. The fastest collision leading to sticking has to be considered apart from the others, though. As can be seen in Fig. \ref{fig:parameterplotwithdata}, the smaller dust aggregate had a mass of only $10^{-6}$ g, leading to a mass ratio of the collision partners of 133. In this case, a small fragment stuck to another particle, violating the assumption of the model that the particles are roughly equal sized. Therefore, we neglect this collision in our further considerations.

From our experiments, we can conclude that the change from sticking to bouncing is not abrupt but is rather characterized by a transition zone where particles stick or bounce with some probability. This will be discussed in detail in Sect. \ref{sec:discussion}, but before we will statistically characterize the collisions in further detail.

\subsection{Impact parameters}\label{sub:collisionparameter}
The impact parameter $b$ describes how centered or off-center a collision is taking place. It is a measure for the vector component of the distance between the centers of mass of the two colliding particles at the impact time perpendicular to their relative velocity vector. By normalizing it with $R = r_1 + r_2$, where $r_1$ and $r_2$ are the radii of the two colliding particles (assuming spherical shape), a value of $b/R = 0$ indicates a perfectly central collision while $b/R = 1$ denotes a grazing collision.

\begin{figure}[!t]
    \includegraphics[width=\columnwidth]{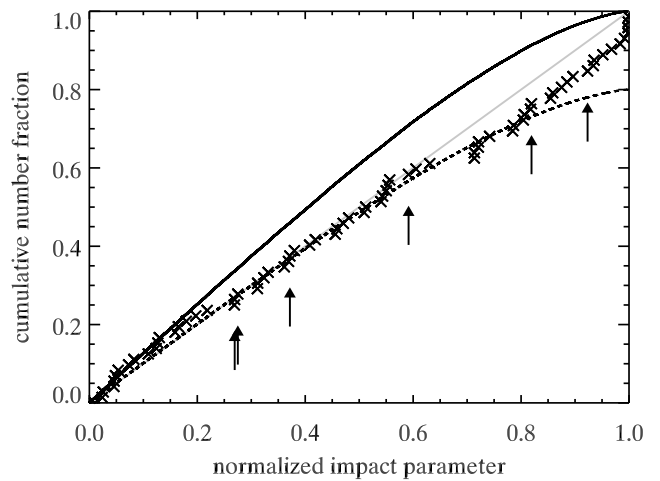}
    \caption{\label{fig:distributioncollisionparameter}The distribution of the impact parameter of the 72 investigated dust-aggregate collisions in the two-dimensional view. The solid gray line shows a random distribution of 2D impact parameters, the black solid line represents the expected distribution for a two-dimensional viewing geometry in a three-dimensional experiment, and the black dashed line is the best fitting distribution taking the collision statistics into account. The six collisions leading to sticking are marked with arrows and appear to be randomly distributed.}
\end{figure}

\begin{figure}[t]
    \includegraphics[width=\columnwidth]{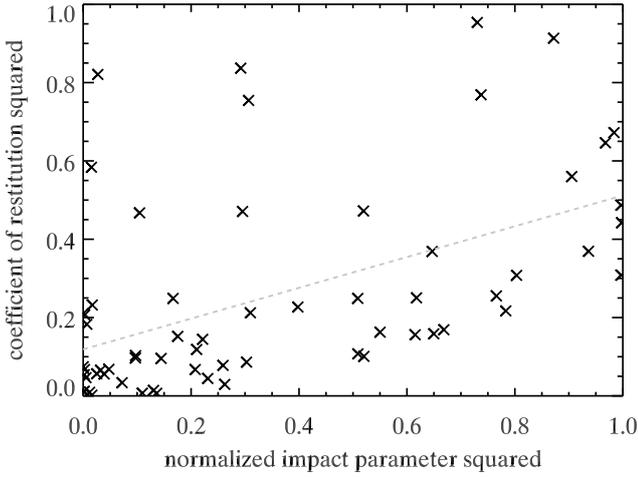}
    \caption{\label{fig:epsoverb}The two-dimensional coefficients of restitution of 66 bouncing collisions between 1 mm dust aggregates show a significant scatter but possess a general increasing trend with increasing impact parameter. Although only weakly correlated for the squared values as plotted here (linear Pearson correlation coefficient of 0.5), a linear fit agrees with the results of \citet{BlumMuench:1993}.}
\end{figure}

The positions of the centers of mass were calculated from the fitted trajectories of the 1 mm dust aggregates before and after each collision. In the cases where the particle tracks after the collision could not be analyzed (i.e. for sticking collisions or in collisions where one dust aggregate was partly covered), the existing data was insufficient to calculate the impact parameter properly. Therefore, we had to discard 31 collisions, leaving a total of 72 collisions including 6 sticking collisions, which we examined individually. Figure \ref{fig:distributioncollisionparameter} shows the distribution of the normalized impact parameter $b/R$, marked by the crosses, in a cumulative diagram. The six sticking collisions are marked by arrows. Within the limits of the small number of events, they can be regarded as being randomly distributed over the impact parameter range, which shows that the impact parameter is not the main driver for the sticking in these collisions. The impact parameters of all analyzed collisions also seem to be randomly distributed (gray line in Figure \ref{fig:distributioncollisionparameter}), but the expectation for the distribution of a two-dimensional projection of impact parameters is rather represented by the black solid line, which was deduced from a simple Monte Carlo simulation of arbitrary collisions. The reason why central collisions are expected to be more dominant in the projection is that the cross section in the direction that is not observed is much bigger than for grazing collisions. From the comparison of our data with the black dashed line, representing the same distribution but with a different normalization, it is obvious that large impact parameters near unity are over-represented. There are several possible reasons for this: \emph{(i)} with the dust aggregates being elliptic rather than perfectly spherical and the normalization of the impact parameter being based on an equivalent spherical radius, impact parameters $b/R > 1$ are possible. Although we do not see any of those, this still leads to an overestimation of the impact parameter. \emph{(ii)} Irregular surfaces may counteract against the argument of decreasing cross section with increasing projected $b / R$ (see above). \emph{(iii)} As shown in Fig. \ref{fig:epsoverb} the coefficient of restitution is generally larger with bigger impact parameter, meaning that less translational energy is lost in the collision and the particles separate faster than they would after a central collision. The less time (i.e. the fewer images) passes after a collision until the particles can be tracked again, the more likely it is that no other dust aggregate has moved to the same area in the meantime and the collision partners can still be observed.

The coefficient of restitution $\varepsilon$ is a measure of the loss of energy in a collision. It is defined as the ratio of the relative velocity of the colliding objects after and before a collision. While $\varepsilon = 0$ denotes a perfectly inelastic collision, a value of  $\varepsilon=1$ means that no energy was lost in the collision, i.e. a perfectly elastic collision. In Fig. \ref{fig:epsoverb} the squared coefficient of restitution of 66 bouncing collisions in which we could follow both particles before and after the impact is plotted over the squared impact parameter. This analysis is limited to the 1 mm dust aggregates because we did not follow the rebound velocities in the experiments with the larger particles. Although our 2D experiment shows a wide scatter in the coefficient of restitution, a general trend of a higher coefficient of restitution with growing impact parameter can be observed. The linear Pearson correlation coefficient of the squared values is 0.5, indicating at least a moderate correlation. \citet{BlumMuench:1993} showed this trend to be linear and derived a value of $\varepsilon^2=0.51$ for the squared coefficient of restitution for a grazing collision (i.e. $b/R = 1$) of spherical particles with rough surfaces that prevent sliding during contact. Using a chi-square fit we obtain values of $\varepsilon^2 (b^2/R^2 = 0) = 0.12 \pm 0.04$ and $\varepsilon^2 (1) = 0.51 \pm 0.12$, in agreement with the analysis of \citet{BlumMuench:1993}, although our particles are not perfectly spherical and are likely to have small differences in porosity and size.

As a reason for the scatter in the coefficient of restitution data, we identify the rough and irregular surface and the unknown rotation of the dust aggregates. As an example, a dust aggregate can store rotational energy in one collision and release it into translational kinetic energy in the next collision. Moreover, the model of \citet{BlumMuench:1993} assumes spherical particles, which yield a linear relation between the impact parameter and the sine of the impact angle. In the first approximation, our particles are represented by ellipsoids, for which the difference in impact angle can easily be of the order of 15° for any given impact parameter. Moreover, the size ratio of our particles can be up to 1:3 in extreme cases, while it is assumed to be unity in the model.

\section{Discussion}\label{sec:discussion}

In this section we will discuss the actual thresholds between sticking and bouncing collisions. In Sect. \ref{sub:probability} we will present our results in terms of a velocity-dependent sticking probability and in Sect. \ref{sub:model} we adapt a theoretical collision model to deduce the velocity threshold below which all dust aggregates stick. In Sect. \ref{sec:dimer} we will discuss a peculiar collision between a single dust aggregate and two dust aggregates sticking together. From this we will get insight into the binding forces between sticking dust aggregates in our experiments. A brief discussion of the implications of these results on the growth model is given in Sect. \ref{sec:consequences}.

\subsection{Sticking probability}\label{sub:probability}

From the results of the experiment with the smaller 1 mm dust aggregates presented in Sect. \ref{sub:collisionresults} it is apparent that sticking of particles also occurs at velocities higher than proposed by the dust-aggregate collision model of \citet{GuettlerEtal:2010}. Although we could observe only seven collisions in which the particles stuck to one another, which is just a small fraction of the total number of observed collisions, there is a clear trend of sticking happening at lower velocities. For the following considerations we discard the single sticking event of the two different-sized dust aggregates. However, this collision does not change the overall picture.

\begin{figure}[t]
    \includegraphics[width=\columnwidth]{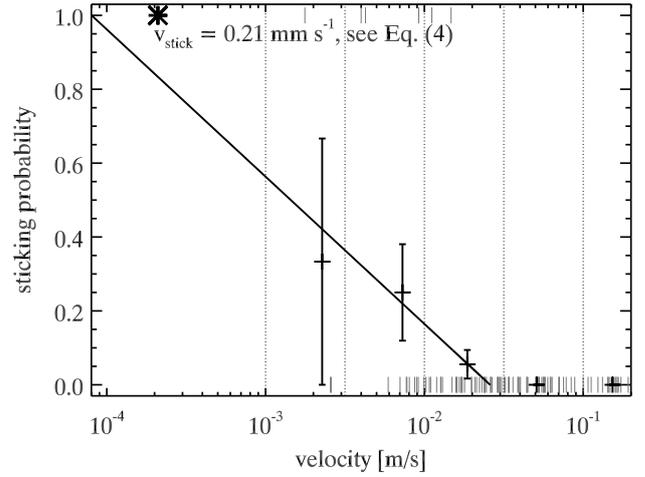}
    \caption{\label{fig:sticking_probability}The transition between sticking and bouncing collisions is not sharp but follows a probability distribution in velocity (solid line). The black crosses denote the fraction of sticking collisions in the given intervals. The vertical ticks at 0 and 1 denote the individual bouncing and sticking collisions, respectively. The asterisk marks the sticking velocity $v_\mathrm{stick} = 0.21$ \mms\ derived in Sect. \ref{sub:model}.}
\end{figure}

As we observed sticking and bouncing in the same range of velocities and impact parameters, we propose that there is no sharp threshold between sticking and bouncing collisions, but rather a smooth transition over roughly two orders of magnitude in velocity. Figure \ref{fig:sticking_probability} shows the probability of a collision leading to sticking for five velocity intervals between $v = 10^{-3}$ \metersecond\ and $v = 0.2$ \metersecond\ (black crosses). The probability was calculated by dividing the number of sticking collisions in the interval by the total number of collisions. Single collisions are indicated by the vertical bars at sticking probabilities 0 (bouncing) and 1 (sticking). The solid line in Fig. \ref{fig:sticking_probability} shows a logarithmic fit
\begin{equation}
    P_\mathrm{stick}(v) = -0.6 - 0.4\cdot \lg \frac{v}{\mathrm{m\,s}^{-1}} \label{eq:stick_prob}
\end{equation}
(for $8\cdot 10^{-5}$ \metersecond $< v < 2.6\cdot 10^{-2}$ \metersecond) to the three values of the sticking probability that are nonzero. Although this probability should be taken with care, it nicely reproduces the 100 \% probability sticking velocity of $2.1\cdot 10^{-4}$ \metersecond, which we will derive in Sect. \ref{sub:model}.

In the collisions between 1.5 mm dust aggregates we did not observe any sticking events. To statistically estimate the number of sticking collisions to be expected for these dust aggregates, we use the velocity of each bouncing collision to calculate the expected sticking probability. Adding up all these probabilities yields an expected number of 2.2 sticking collisions in these experiments, which should be observed if the 1.5 mm dust aggregates behaved similar to the 1 mm dust aggregates. Correcting the velocity by a factor of $\left( m_1 / m_2 \right) ^{- \frac{5}{18}} = 1.7$, with $m_1$ and $m_2$ being the average mass of the smaller collision partner of the 1 and 1.5 mm dust aggregates as displayed in Fig. \ref{fig:parameterplotwithdata}, respectively, leads to an expected 1.2 sticking collisions. This correction takes a mass dependency of the sticking probability into account following the model by \citeauthor{GuettlerEtal:2010} and Sect. \ref{sub:model}. However, we do not see any sticking collision for the 1.5 mm dust aggregates. Using Eq. (\ref{eq:stick_prob}) we calculated the probability of observing no sticking collision for the given number of collisions with their respective velocities to be 8\%. If we correct the velocities as above to account for the higher masses, we get a probability of 25\%\ of not seeing any sticking collision. Therefore, our observations do not contradict the expected result.

\subsection{A model for the sticking velocity}\label{sub:model}
\begin{figure}[t]
    \includegraphics[width=\columnwidth]{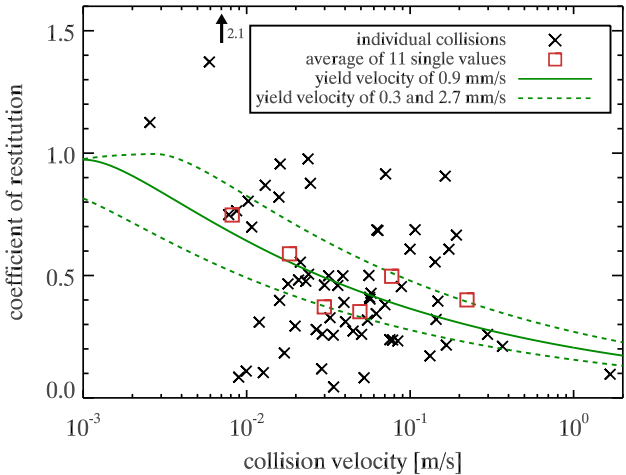}
    \caption{(color online) Our data for the coefficient of restitution as a function of collision velocity for all impact parameters. The red squares are averages over 11 single values, which are compared to the model by \citet{ThorntonNing:1998}, denoted by the green solid and dashed curves.}
    \label{fig:epsoverv}
\end{figure}

From the coefficients of restitution presented in Sect. \ref{sub:collisionparameter} we can learn important dynamic material parameters of the millimeter-sized dust aggregates. \citet{ThorntonNing:1998} developed a model for the velocity dependence of the coefficient of restitution in collisions between elastic-plastic adhesive spheres, described by their Equations (80) and (81). In  Fig. \ref{fig:epsoverv} we present our coefficients of restitution as a function of the impact velocity (black crosses) and the mean values of 11 consecutive data points (red squares; arithmetic mean in $\varepsilon$, geometric mean in $v$). There is a lot of scatter that we attribute to the distribution of impact parameters (see Figs. \ref{fig:distributioncollisionparameter} and \ref{fig:epsoverb}), irregular surfaces, and rotation. Values above $\varepsilon = 1$ are possible if dust aggregates are rotating before the collision and transform rotational energy into translational energy; one collision even had a coefficient of restitution of $\varepsilon = 2.1$. In the averaged data, however, a clear trend of decreasing coefficient of restitution with increasing velocity can be found, which can be compared to the model by \citeauthor{ThorntonNing:1998}. Here, we ignore that the model was developed for central collisions while our collision parameters are randomly distributed. We will comment on that point later, but if we keep this constraint in mind, we will arrive at interesting results. The only free parameters in the model are the sticking velocity $v_\mathrm{stick}$ and the yield velocity $v_\mathrm{yield}$ of the dust aggregates. The yield velocity is the minimal velocity for which the so called yield pressure $p_\mathrm{yield}$ is achieved at the contact surface, which marks the onset of plastic deformation. The presented curves in Fig. \ref{fig:epsoverv} are mainly determined by the yield velocity, the sticking velocity defines a cut-off at low velocities, which occurs in this plot at velocities lower than $10^{-3}$ \metersecond. Our mean data (red squares) agrees with the model for $v_\mathrm{yield} = 0.9^{+1.8}_{-0.6}$ \mms\ (green solid and dashed curves). The relation between yield pressure and yield velocity as given by \citeauthor{ThorntonNing:1998} follows from their Equation (10) as
\begin{equation}
    v_\mathrm{yield} = 6.4\sqrt{\frac{p_\mathrm{yield}^5 r^3}{E^4 m}}, \label{eq:yield_velocity}
\end{equation}
where $r$, $m$, and $E$ are the radii, masses and Young's moduli of similar particles, respectively.

\begin{figure}[t]
    \includegraphics[width=\columnwidth,angle=180]{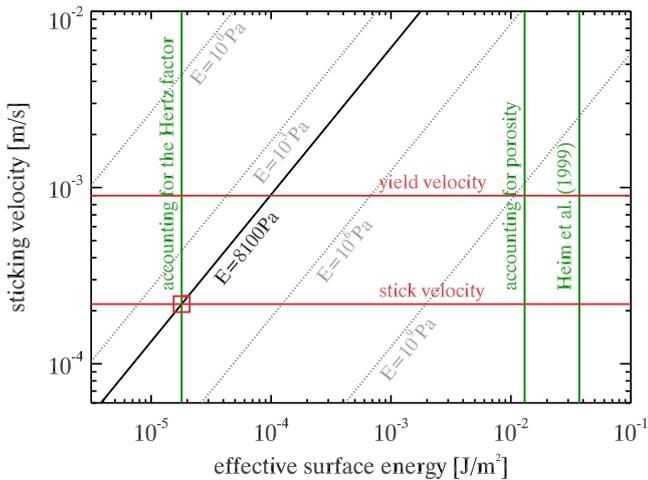}
    \caption{(color online) Graphical illustration of the sticking velocity following Equation (\ref{eq:stick_velocity_model}) for a range of surface energies and Young's moduli. Our best values for the surface energy ($1.8 \cdot 10^{-5}\ \mathrm{J\,m^{-2}}$) and Young's modulus (8100 Pa) result in a sticking velocity of 0.21 \mms.}
    \label{fig:TN_conclusion}
\end{figure}

As a measure for the yield pressure, we use the static compression measurements by \citet{BlumEtal:2006}, who used dust aggregates from the same dust as in our experiments. For a pressure of 200~Pa, the dust-aggregate material started to yield to the applied pressure, which we therefore take as the yield pressure. If we assume dust aggregates with $r=0.5$~mm and $\varrho=910$~\density, we can use Equation (\ref{eq:yield_velocity}) to calculate Young's modulus of our dust aggregates to be $E=8100$~Pa, which is about an order of magnitude higher than earlier assumptions \citep[e.g.][and others]{GuettlerEtal:2009}. The sticking velocity in the model of \citeauthor{ThorntonNing:1998} (their Equation (54)) is
\begin{equation}
    v_\mathrm{stick} = 4.2\left(\frac{\gamma_\mathrm{eff}^5r^4}{m^3E^2}\right)^{1/6} \ ,\label{eq:stick_velocity_model}
\end{equation}
where $\gamma_\mathrm{eff}$ is the surface energy of both surfaces in contact, which we call effective surface energy. This surface energy is a complicated parameter in this context as we have two very complex surfaces in contact. The model by \citeauthor{ThorntonNing:1998} assumes a smooth contact surface that uniformly contributes to the contact energy. In our case, we use the \emph{effective} surface energy as a combination of the known surface energy (for a smooth surface), the dust aggregate porosity, and the Hertz factor (the ratio between contact surface and cross section of two monomer grains) and apply
\begin{equation}
    \gamma_\mathrm{eff} = \gamma \cdot \phi \cdot \frac{a^2}{a_0^2} \ . \label{eq:surface_energy}
\end{equation}
The first term on the rhs. of Equation (\ref{eq:surface_energy}) is the surface energy of the \sio\ material measured for micrometer-sized monomer grains by \citet{HeimEtal:1999} to be $\gamma=0.037\ \mathrm{J\,m^{-2}}$ (energy of both surfaces in contact). The second term is the volume filling factor, which we use as a surface filling factor of $\phi = 0.35$ and that accounts for the dust-aggregate porosity. Still, two grains in contact do not have a contact which is as large as their cross section but smaller than this, due to the Hertz factor. The Hertz factor, $(a/a_0)^2$, is the ratio between the contact cross section of two spherical dust monomers $\pi a^2$ and the monomer cross section $\pi a_0^2$. We approximate the contact geometry of our irregular grains by that of spherical grains (the geometric mean diameter of our grains is 1.0~\mum, Young's modulus is 41~GPa, and Poisson's ratio is $\nu = 0.17$) to calculate the Hertz factor. For the contact radius $a$ we assume spheres sticking to each other only due to their adhesion. The contact radius is then given by \citet[their Equation (20)]{JohnsonEtal:1971} as
\begin{equation}
    a = \left(\frac{9 \pi \gamma (1-\nu^2) a_0^2}{E_0}\right)^{1/3} \ ,
\end{equation}
and the Hertz factor is consequently
\begin{equation}
    \frac{a^2}{a_0^2} = \left(\frac{9 \pi \gamma (1-\nu^2)}{a_0 E_0}\right)^{2/3} \ .
\end{equation}
For the grains assumed here, this results in $a^2/a_0^2 = 1.4 \cdot 10^{-3}$ and the effective surface energy including both geometric effects is as low as $\gamma_\mathrm{eff} = 1.8 \cdot 10^{-5}\ \mathrm{J\,m^{-2}}$. Figure \ref{fig:TN_conclusion} graphically illustrates the results of the above considerations: based on Equation (\ref{eq:stick_velocity_model}) we intend to deduce the sticking velocity for a material with sparsely known Young's modulus and an effective surface energy which largely depends on geometry. The best estimate for Young's modulus is 8100 Pa (see above), which is the black, diagonal line. The surface energies are denoted by the green vertical lines and the best value for the sticking velocity is at 0.21 \mms, which is represented by the intersection of the left vertical line (best value for the effective surface energy) and the black diagonal line (best value for the dust-aggregates' Young's modulus). A wide range of Young's moduli from the softest up to solid materials (gray lines) as well as the surface energies with the three corrections (green lines) are plotted for comparison.

The prediction for the sticking velocity for dust aggregates of 0.1 mg mass was approx. $10^{-3}$ \metersecond\ in the model of \citet{GuettlerEtal:2010}. The assumption behind that was that 95\% of the kinetic energy in a collision is instantly dissipated due to plastic deformation ($\varepsilon \simeq 0.2$) and the remaining energy is balanced by the contact energy of the grains. However, a successful adaptation of the model of \citeauthor{ThorntonNing:1998} implies a sticking velocity in a more elastic regime, where sticking is dominated by adhesion rather than by plastic energy dissipation, i.e. the coefficient of restitution is significantly larger than $\varepsilon=0.2$. This is clear from the mean values in Fig. \ref{fig:epsoverv} and also indicated by the green solid line. A sticking velocity of $10^{-3}$ \metersecond\ does not directly contradict our data but should instead be interpreted as a 50 \% sticking probability.

Recapitulated, the main simplifications were the following: \emph{(i)} the average coefficient of restitution presented in Fig. \ref{fig:epsoverv} is also averaged over impact parameters (i.e., $b/R \approx 0.5$), which results in a systematic shift to higher values of $\varepsilon$ (cf. Fig \ref{fig:epsoverb}). Due to projection effects, the unobserved component of the impact parameter makes this correction even greater. \emph{(ii)} The measurement of the coefficients of restitution is also constricted due to the projected view. It is reasonable to assume that collisions predominantly occur with major velocity components in the (observable) direction of the excitation (see \ref{app:projection}). After the collision, the directions of the velocities are randomized due to the non-central collisions and irregular surfaces. This means that the velocity after the collision is more underestimated than the velocity before, which leads to a systematic underestimation for the coefficient of restitution. Qualitatively, this compensates the first shortcoming, while a deeper analysis is complicated. \emph{(iii)} The effective surface energy relies on the Hertz factor, for which we made an assumption of spherical, monodisperse grains (logarithmic mean grain size) as this factor is impossible to describe analytically for irregular grains. \emph{(iv)} The correction for porosity in Equation (\ref{eq:surface_energy}) assumes two porous surfaces, with only Hertzian deformation. Porous surfaces are however rough and might interlink, which increases the effective number of grains in contact.

The sticking velocity given by \citet{GuettlerEtal:2010} depends on the projectile mass as $v\propto m^{-5/18}$. This is consistent with the model of \citeauthor{ThorntonNing:1998} and also with our data for the larger dust aggregates. We therefore regard the current threshold of \citeauthor{GuettlerEtal:2010}
\begin{equation}
    v_\mathrm{stick} = \left(\frac{m}{\widetilde{m}}\right)^{-5/18}\ \mathrm{m\,s^{-1}}
    \approx \left(\frac{m}{0.1\ \mathrm{mg}}\right)^{-5/18}\ \mathrm{mm\,s^{-1}}
    \label{eq:v_stick}
\end{equation}
with $\widetilde{m} = 1.1 \cdot 10^{-18}$ kg as a good estimate for the velocity of a 50 \% sticking probability. In Sect. \ref{sub:probability} we showed that the thresholds for perfect sticking and perfect bouncing should be given by $\widetilde{m} = 1.9 \cdot 10^{-22}$ kg (sticking) and $\widetilde{m} = 2.1 \cdot 10^{-13}$ kg (bouncing).

For particles with a given mass, this leads to a transition in velocity with a width of a factor of 325. The physical reason for this is not easy to deduce for these realistic particles, which feature a rough, irregular surface. The distribution of impact parameters is expected to be of major importance as it also has a large influence on the coefficient of restitution (cf. Fig. \ref{fig:epsoverb}). However, for a central collision we also expect a transition region from perfect sticking to perfect bouncing if we consider a distribution of dust-aggregate sizes from 0.5 to 1.5 mm with an ellipsoidal shape. According to Eq. (\ref{eq:stick_velocity_model}) smaller particles are sticking at higher velocities than larger ones. If those particles are oblate ellipsoids (aspect ratio 1:1.7:1.7, disk shaped) colliding with their flat sides, the surface curvature at the contact site is enhanced by a factor of 1.7. For larger particles, we see from Eq. (\ref{eq:stick_velocity_model}) that the sticking threshold velocity is generally expected to be smaller. If we furthermore consider two prolate ellipsoids (aspect ratio 1:1:1.7, rod shaped) in the extreme case, which collide with the long axes aligned, the surface curvature is reduced by the factor 1.7. We can rewrite Eq. (\ref{eq:stick_velocity_model}) as $v_\mathrm{stick} \propto R^{-3/2} R'^{2/3}$, where the first term represents the mass dependence in Eq. (\ref{eq:stick_velocity_model}), i.e., the inertia of the dust aggregates, and $R'$ in the second term describes the influence of the surface curvature. The width of the sticking transition from the two effects considered here is thus
\begin{equation}
    \frac{v_\mathrm{s}}{v_\mathrm{l}} = \left(\frac{R_\mathrm{s}}{R_\mathrm{l}}\right)^{-3/2}\left(\frac{1.7 R_\mathrm{s}}{R_\mathrm{l}/1.7}\right)^{2/3}\ ,
\end{equation}
where the indices s and l represent the small and large radii. This ratio can explain a width in velocity of a factor 5 for the transition from sticking to bouncing. This can be significantly larger if we also consider local changes in the radius of curvature (e.g., small bumps and dimples), which can be much more diverse.

\subsection{Contact strength of sticking dust aggregates}\label{sec:dimer}
As mentioned before, seven collisions led to sticking of the dust aggregates. The qualitative impression in the movie sequences (see online material) is that the bonding between sticking particles appears rather weak. One of the dimers even collides with a glass wall (with unknown velocity), and the two particles separate again. However, one of the other dust-aggregate dimers collides with a single dust aggregate, and we have a good observation angle. This allows us to estimate the strength of the contact and obtain a quantitative estimate where the qualitative impression might be deceptive. We will first calculate a minimum shear strength of the connecting neck and then compare the kinetic energy in the collision with the adhesive energy in this neck.

\begin{figure}[t]
    \includegraphics[width=\columnwidth]{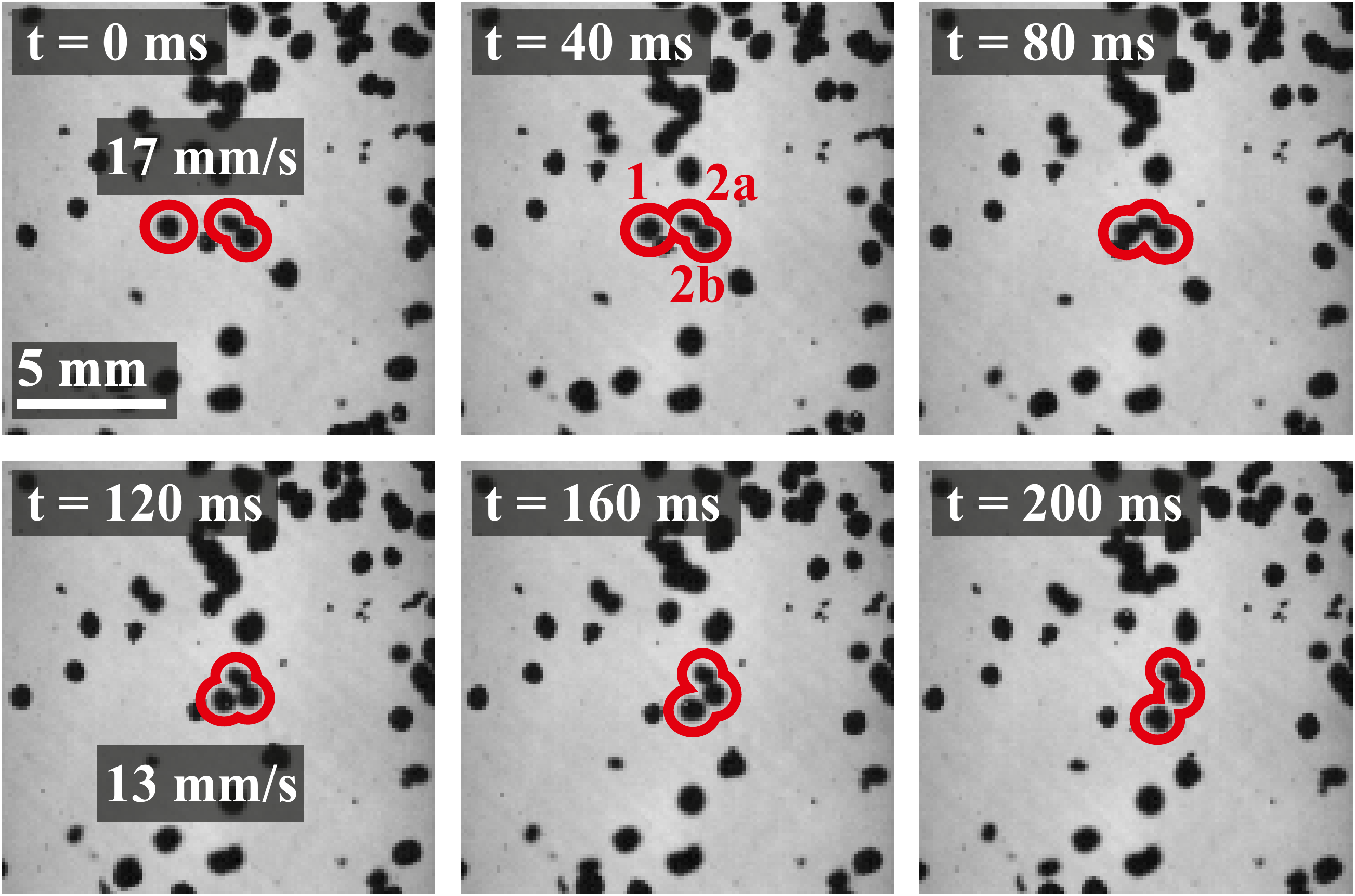}
    \caption{\label{fig:dimer_sequence}(color online) The image sequence shows a dust-aggregate dimer being hit by a single dust aggregate. First, the single dust aggregate collides with the upper dust aggregate of the dimer, which then rotates around the lower one. After this, the lower particle is hit as well and the dimer is accelerated to the right, but still intact. A movie of this collision can be found in the online version of this article.}
\end{figure}

The image sequence of this interesting collision is presented in Fig. \ref{fig:dimer_sequence}. The dimer dust aggregate (particle 2) is nearly at rest and the single dust aggregate (particle 1) comes from the left colliding with the upper dust aggregate of the dimer (particle 2a) at a  collision velocity of $1.7\cdot 10^{-2}$ \metersecond. The velocity vector is nearly perpendicular to the axis of the dimer dust aggregate. Particle 2a is hit, accelerated to the right, and then rotates around the common center with particle 2b. With a relative velocity of $1.3\cdot 10^{-2}$ \metersecond, particle 1 then collides with particle 2b of the dimer, causing the dimer to stop its rotation and move to the right from the point of collision. The two particles of the dust-aggregate dimer are still attached to each other and can be followed until the very end of the experiment, 2.8 seconds after $t = 0$ s in Fig. \ref{fig:dimer_sequence}, during which the particles clearly move together.

If the collision vector is perpendicular to the dimer axis, particle 2a might be sheared off if the momentum of the single dust aggregate is sufficiently large. We will calculate the shear strength of the connecting neck between particles 2a and 2b, for which this could happen, which is then a lower estimate for the actual shear strength of the dust-aggregate contact. If the neck has a cross sectional area $A$ and particle 1 is pushing with a constant force $F$, the shear strength must be $S > F/A$ to withstand. For the force we make an approximation to distribute the momentum of particle 1 ($m_1 \cdot v_1$) equally over the impact time and assume a contact time of $\tau= 5$ ms as discussed by \citet{WeidlingEtal:2009}. With this we get
\begin{equation}
    F = \frac{m_1 v_1 (1+\varepsilon)}{\tau} \ .
\end{equation}
For the contact area, we assume that the dimer was grown in a collision with elastic deformation and take the Hertzian contact area given by \citep{Johnson:1985} as
\begin{equation}
    A_\mathrm{Hertz} = \pi\left(\frac{15}{8} \frac{m^* r^{*2}v^2}{E}\right)^{2/5} \ ,
\end{equation}
where $m^*$ and $r^*$ are the reduced mass and radius of particles 2a and 2b, $v = 4.3\cdot 10^{-3}$ \metersecond\ is the velocity of the collision from which the dimer had grown, and $E=8100$ Pa is Young's modulus (see Sect. \ref{sub:model}). For the dust aggregates with radii of $r_1 = 0.33$ mm, $r_{2\mathrm{a}} = 0.25$ mm, and $r_{2\mathrm{b}} = 0.30$ mm, this yields a contact area of $A_\mathrm{Hertz} = 9.3 \cdot 10^{-10}\ \mathrm{m^2}$ and a shear strength of $S>640$ Pa. This is a reasonable value and in agreement with compressive and tensile strengths in the range of 1000 Pa \citep{GuettlerEtal:2009}.

We can also compare the kinetic energy in the collision which is not dissipated due to plastic deformation (cf. Sect. \ref{sub:model}) with the adhesive energy in the contact. For the kinetic energy we chose to take the reduced mass of only particles 1 and 2a as particle 2b appears to be nearly unaffected by the first collision. The adhesive energy is $N \cdot E_\mathrm{break}$ where $N$ is the number of sticking contacts and $E_\mathrm{break} = 3.4\cdot 10^{-16}$ J is the contact energy between two grains with radius $a_0=0.5$ \mum\ \citep{PoppeEtal:2000a,BlumWurm:2000}. With $N > \varepsilon^2 E_\mathrm{kin}/E_\mathrm{break}$ we get $N>1700$ contacts, which represents a contact area of
\begin{equation}
    A_\mathrm{adh} = \frac{N \pi a_0^2}{\phi} \ .
\end{equation}
With this, we get a contact area between the two dust aggregates $A_\mathrm{adh} > 3.8 \cdot 10^{-9}\ \mathrm{m^2}$, which is significantly larger than calculated above. A more intuitive parameter is the contact radius which is $a_\mathrm{Hertz}=17$ \mum\ and $a_\mathrm{adh}=35$ \mum, respectively.

We have to note that the energy consideration has the problem that it assumes that all the kinetic energy left after plastic deformation goes into breaking the contacts. For the shear strength considerations, we showed that the velocity to break a dimer is proportional to the shear strength, which may well be a factor of two higher than the minimum calculated here. This yields a breakup velocity around 0.3 \metersecond, which is higher than the highest sticking velocity found in our experiments (see, e.g., Fig. \ref{fig:sticking_probability}). Thus, if we find conditions that favor sticking in a quiescent region in a protoplanetary disk, sticking will continue and further collisions at the same velocity will not break contacts again.

\subsection{Consequences for the protoplanetary dust growth}\label{sec:consequences}
The consequences of our experiments for the growth of protoplanetary dust aggregates are not easy to predict. The model of \citet{GuettlerEtal:2010} is highly non-linear and small changes can have significant impact. However, it is possible to give an indication about the mass of the dust aggregates \citep[relative to the mass computed by][]{ZsomEtal:2010, ZsomEtal:preprint} at the time when the dust aggregates cross the boundary between sticking and bouncing for the first time.

In Sect. \ref{sub:model} we showed that the sticking threshold velocity scales with mass according to Eq. \ref{eq:v_stick}. To determine the ratio of two masses for two different scaling factors of the threshold line $\widetilde{m}_\mathrm{old}$ \citep{GuettlerEtal:2010} and $\widetilde{m}_\mathrm{new}$ (this work), we can write
\begin{equation}
	\frac{m_\mathrm{new}}{m_\mathrm{old}} = \left(\frac{v_\mathrm{new}}{v_\mathrm{old}}\right)^{-18/5} \frac{\widetilde{m}_\mathrm{new}}{\widetilde{m}_\mathrm{old}} \ ,\label{eq:velocity_mass_scaling}
\end{equation}
where the first term on the rhs. denotes the ratio of typical collision velocities for the dust aggregates with the respective indices. If the relative velocity for the dust aggregates in the protoplanetary disk was not dependent on their size, this term would be unity and we just get the ratio of the two scaling factors. However, we know that the relative velocities are increasing with the dust aggregates' masses as $v \propto m^{1/3}$ under the conditions considered here \citep{WeidenschillingCuzzi:1993}, and thus we get
\begin{equation}
	\frac{v_\mathrm{new}}{v_\mathrm{old}} = \left(\frac{m_\mathrm{new}}{m_\mathrm{old}}\right)^{1/3} \ .
\end{equation}
Using this for Eq. \ref{eq:velocity_mass_scaling} we arrive at
\begin{equation} \frac{m_\mathrm{new}}{m_\mathrm{old}}=\left(\frac{\widetilde{m}_\mathrm{new}}{\widetilde{m}_\mathrm{old}}\right)^{5/11} \ ,
\end{equation}
which yields a significant factor of 250.

An exact mass for the largest dust aggregates that can directly grow shall not be given here, as this number could easily be misinterpreted. It must be stated that this is the enhanced growth only for one certain time in the evolution of protoplanetary dust. After this, new non-linear effects could show up in the model of \citet{GuettlerEtal:2010} and \citet{ZsomEtal:2010} so that the overall growth might be enhanced in the same way but could as well stall at the same mass as before. An interesting aspect is, however, that the so-called 'bouncing barrier' \citep{ZsomEtal:2010} is becoming smaller: growth is enhanced and also the fragmentation threshold was recently revised \citep{BeitzEtal:2011a} and found to be at smaller velocities for large (centimeter) dust aggregates. Thus, it might be easier to trigger fragmentation, widen the dust-aggregate size distribution, and by this open a new path for the further protoplanetary dust-aggregate growth.

\section{Conclusions}\label{sec:conclusion}
With a novel experimental setup, we conducted microgravity experiments with a multitude of porous dust aggregates of different sizes and analyzed free collisions between them. We found a total of 125 collisions with relative velocities between 2 \mms\ and 1.7 \metersecond, where 22 collisions happened between 1.0 to 2.0 mm-sized dust aggregates that all resulted in bouncing and 103 collisions between 0.5 to 1.5 mm-sized dust aggregates (see Fig. \ref{fig:parameterplotwithdata}). Of those 103 collisions we observed sticking in seven cases, bouncing in 95 collisions and once one of the collision partners fragmented. This was the first time that sticking of millimeter-sized dust aggregates was observed under controlled experimental conditions.

The sticking collisions found in the experiment scatter over a wide range of velocities, all of which are higher than predicted by the model of \citet{GuettlerEtal:2010}. Therefore, we suggest to implement a transition zone from sticking to bouncing collisions with a velocity-dependent probability for the collision partners to stick to each other. The threshold line given by \citeauthor{GuettlerEtal:2010} is consistent with a 50 \% sticking probability in our experiments, while the velocity thresholds for a 100\% and 0\% sticking probability are given by our Equation (\ref{eq:v_stick}) with $\widetilde{m} = 1.9 \cdot 10^{-22}$ kg (100\% sticking probability) and $\widetilde{m} = 2.1 \cdot 10^{-13}$ kg (0\% sticking probability).

To determine a general description for the fastest velocity that always leads to sticking, we adapted the collision model by \citet{ThorntonNing:1998} in Sect. \ref{sub:model}. We determined Young's modulus of our dust aggregates to be $E = 8100$ Pa by deriving a yield velocity of $v_\mathrm{yield} \approx 9\cdot 10^{-4}$~\metersecond\ from the decline of the coefficient of restitution with increasing collision velocity. By accounting for porosity of the dust aggregates and the Hertz factor we received a term for the effective surface energy of our dust aggregates, which we use to formulate an analytic term for the sticking velocity as given by Equation (\ref{eq:stick_velocity_model}).

Furthermore, we could show that the contact between sticking dust aggregates of millimeter size is sturdy enough to let the dimer survive collisions with single dust aggregates if the velocity is low enough to provide a chance for sticking.

\subsection*{Acknowledgements}
We are grateful to the Deutsches Zentrum für Luft- und Raumfahrt (DLR) for funding the experiments and drop tower campaigns under grant 50WM0936. C.G. was funded by the Deutsche Forschungsgemeinschaft within the Forschergruppe 759 ``The Formation of Planets: The Critical First Growth Phase'' under grant Bl 298/14-1.

\bibliographystyle{apalike}
\bibliography{literatur}

\begin{appendix}

\section{Gas drag on the dust aggregates}\label{app:gasdrag}
Particles moving in vacuum under microgravity conditions are expected to do so on linear trajectories with constant velocity. However, disturbances like static charging of the particles or drag of the residual gas may alter this behavior. While we can rule out the former for those 1 mm dust aggregates undergoing collisions by having checked their trajectories for parabolic components we have to address the latter more thoroughly, as gas drag slows down the particles exponentially even without collisions. A measure for the importance of gas drag is the stopping time $\tau_\mathrm{f}$, after which a particle's velocity has decreased to $1/e$ of its starting value. With a gas pressure of $p = 10$ Pa the mean free path of the gas is $l_g \approx 1$ mm, which is comparable to the radius $r$ of our particles. Therefore, we calculate the friction time both for the Stokes and the Epstein regime after \citet[their Sect. 5.1]{Blum:2006}. They give the stopping time as
\begin{equation}
    \tau_F =
    \begin{cases}
        \frac{2 r^2 \rho_{\mathrm{b}} \phi}{9 \mu_{\mathrm{g}}} &: l_{\mathrm{g}} \ll r \mathrm{\ (Stokes)} \\
        \frac{r \rho_{\mathrm{b}} \phi}{\rho_{\mathrm{g}} \overline{v}_{\mathrm{g}}} &: l_{\mathrm{g}} \gg r \mathrm{\ (Epstein)}
    \end{cases}
    \label{eq:tauf}
\end{equation}
with $\rho_{\mathrm{b}} = 2600$ \density\ \citep[Table 1]{BlumEtal:2006}, $\rho_{\mathrm{g}} = 1.2\ \mathrm{kg\, m^{-3}} \cdot (p/10^5\ \mathrm{Pa})$, $\phi = 0.347$, $\mu_\mathrm{g} = 17.2\ \mathrm{\mu Pa\, s}$ and $\overline{v}_{\mathrm{g}} = 655$ \metersecond\ being the bulk density of the dust particles, the gas density, the filling factor of the particles, the viscosity and the thermal velocity of the gas, respectively. With particle radii between $r_\mathrm{min} = 0.25$ mm and $r_\mathrm{max} = 0.75$ mm, this corresponds to stopping times between 0.7 and 9.3 s, depending on the particle size and the friction regime. Although both times are of the order of the experiment duration, most trajectories are rather short due to inter-particle and particle-wall collisions and the exponential function can be approximated linearly over short sections (here 60 data points, i.e. 120 ms, were used at most). The frictional velocity decay is thus between $\Delta v/v = 1 - \exp \left( -0.12\mathrm{\ s}/0.7\mathrm{\ s} \right) = 0.16$ and $\Delta v/v = 1 - \exp \left( -0.12\mathrm{\ s}/9.3\mathrm{\ s} \right) = 0.01$ for tracks that have the maximum length. Most of the tracks were shorter than that, reducing the frictional decay even more.

\section{Projection effects}\label{app:projection}

\begin{figure}[t]
    \centering
    \includegraphics[width=\columnwidth]{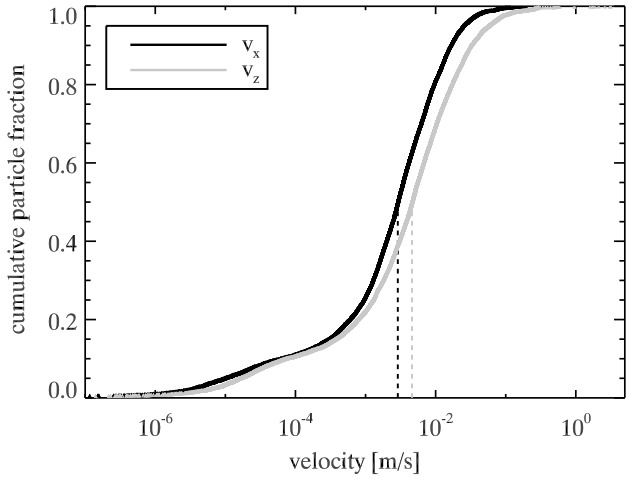}
    \caption{\label{fig:velocitiescumulative}The velocities in the $x$- and $z$-direction were derived directly from the fits. The velocities in the shaking direction are higher than perpendicular to it. The median velocities are $v_x = 2.9$ \mms and $v_z=4.6$ \mms.}
\end{figure}

In order to investigate the error we introduce by using just one projection and not the three-dimensional data we compared the components of the particle velocities. The $z$-component of the velocity, pointing parallel to the shaking movement, is the most prominent term. While in the other directions the particles reach kinetic equilibrium, the shaking induces additional energy in the $z$-direction. Figure \ref{fig:velocitiescumulative} shows the fraction of the particles with a certain velocity or smaller throughout the whole experiment. It can be seen that the particles on average are faster in the $z$- than in the $x$-direction. The median velocity in the $x$-direction is $2.9\cdot 10^{-3}$ \metersecond, while it is $4.6\cdot 10^{-3}$ \metersecond\ in the $z$-direction. We do not expect the velocities in the $y$-direction to differ from those in the $x$-direction, so that the average difference between projected and total velocity is $\sqrt{4.6^2 + 2\cdot 2.9^2} / \sqrt{4.6^2 +  2.9^2} -1 = 0.13$. This would shift the results a little, but not change the overall picture.

\end{appendix}

\end{document}